# Bioremediation of Contaminated Soil with Crude Oil Using Consortium of bacteria

Ahmed Mohamed Taher [a] and Ibrahim Omar Saeed [a, *]

[a, b] Department of Biology, College of Science, University of Tikrit / Salah Al-Din_ Iraq

ABSTRACT

This study included isolate petroleum hydrocarbons degradable bacteria and develops a consortium or a mixture of bacteria with high biodegradation capabilities which can be used in biological treatment units of the contaminated soils before release. In this study ten bacterial strains were isolated from soils contaminated with crude oil, by primary and secondary screening, and by using the sterile saline solution with 1% of crude oil, five bacterial isolates that can degrade oil were identified. The bacterial isolates were isolated from polluted soils with crude oil, the samples were collected from different areas of the Baiji refinery, and samples of heavy crude oil extracted from Qayyarah fields were used in designing the biodegradation experiments, the five isolates were diagnosed based on phenotypic, culture and biochemical characterizes, and detection of the gene sequence of bacterial isolates (16SrRNA). The (16SrRNA) gene sequence of the bacterial isolates was recorded under the accession numbers LC596402, LC596403, LC596406, LC596404, LC596405 for the genus (AM-I-1, AM-I-2, AM-I-5) and the species (AM-I-3, AM-I-4) that following Bacillus Sp respectively, in the NCBI's GenBank, the efficiency of the five isolates were examined for utilizing petroleum hydrocarbons using a sterile mineral salt medium MS supplemented with crude oil as the sole source of carbon with different concentrations, the results showed that the decomposition of petroleum hydrocarbons at the concentrations (0.5, 1, 1.5, 2, 3) % reached (57.32, 69.36, 63.71, 75.20, 68.60) %, respectively, for mixed isolates, depending on the results of the gas chromatography (GC) analysis.

## 1. Introduction

Petroleum hydrocarbon continues to be used as the main supply of energy, accidental spills and leak occur frequently during the transport, exploration, refining, and storage of petroleum and petroleum products, making petroleum an essential global environmental pollutant, Crude oil refineries and petrochemical industries are very important for national development and improvement of the national economy. There was a great demand for crude oil as a main source of energy in civilized societies, which led to a significant increase in the production of crude oil, which is of great value in the local and global market, and this is what He encouraged the establishment of more refineries and petrochemical factories, and the search and exploration for more sources of crude oil to the point of ignoring the result of these activities on the environment and human life.[1]

Since hydrocarbons can move in food chains, treating hydrocarbons polluting the environment is one of the important goals to protect the biosphere from damage to these compounds that affect most living organisms, including humans. Chemical treatment methods dispersed and emulsified hydrocarbon compounds in the event of contamination of aquatic environments, as for the biological methods that are used in various polluted environments, they include the use of microorganisms and are called biodegradation, the biological processes, in general, are environmentally friendly and cost-effective, as they are easy to design and apply; as such they are more appropriate to the public, Most of these organisms that decompose crude oil and hydrocarbons are bacteria that have developed their enzyme systems and these enzymes enable them to consume these compounds.[2] This study came to isolate effective crude oil-degrading bacteria from polluted areas and develop a consortium of bacteria.

## 2. Materials and methods

### 2.1. Sample collection

Soil samples were collected randomly from five different areas of the Baiji and Qayyarah Refineries, and these areas were contaminated with crude oil spills. The samples were taken from (10-20) cm depth cm by a spa Chula and then placed in sterile polyethylene bags and transferred to the laboratory and kept at a temperature of 4 ˚C until use.

The crude oil samples were collected from the Qayyarah fields before conducting transactions on them and placed in sterile and opaque bottles and kept at 4 ˚C until use.

### 2.2 Bacteria isolation

One gram from each of the soil samples was individually suspended and serially diluted in 9 mL distilled water, $10^{-2}$, $10^{-3}$ and $10^{-4}$ dilutions were plated on nutrient agar plates through spread plate technique and incubated for 24 hours at 37 ˚C to obtain soil isolates. after which the isolates were purified and kept at 4 ˚C until use, bacterial colonies were identified by morphological, cultural, and biochemical characteristics.[3]

### 2.3 Preparation of MS broth (culture media)

Mineral salts media was prepared by first dissolving $(NH_4)_2SO_4$ (1.0 g/L), $CaCl_2$ (0.02 g/L), $MgSO_4 \cdot 7H_2O$ (0.2g/L), $K_2HPO_4$ (1.0g/L) and $KH_2PO_4$ (1.0 g/L) in distilled water. $FeCl_3$ (0.05 g/L) was dissolved separately in distilled water and added to the media. The final volume and pH of the MS broth were adjusted to 7.0 with $K_2HPO_4$. The mixture was then heat stirred in a hotplate and autoclaved.[4]

### 2.4 Primary screening

To prevent the growth of the original inoculum, MS agar plates were used instead of Nutrient agar plates, 100 µL of the culture was plated into MS agar plate by spread plate technique. The MS agar media had the following composition: $NH_4SO_4$ (1.0 g/L), $KH_2PO_4$ (1.0 g/L), $K_2HPO_4$ (1.0 g/L), $FeCl_3$ (0.05 g/L), $MgSO_4 \cdot 7H_2O$ (0.2 g/L), $CaCl_2$ (0.02 g/L) and 10 g/L agar-agar with pH adjusted to 7 petroleum either solution (50 µL) was next spread once moisture had evaporated from the first spread. The plates were incubated at 30 °C in an upright position to prevent dripping of the oil. After 7 days of incubation, the isolates that gave a larger diameter were selected.[5]

## 2.5 Secondary screening

The experiment was designed by culturing the selected isolates in 50 mL MS media (pH 7.0), supplemented with 1% crude oil the culture flask was incubated in the shaker at 150 rpm (30 °C) for 12 days, cell turbidity was measured in 600 nm with a spectrophotometer.[6]

## 2.6 Growth efficiency for mix isolates bacteria in crude oil as a carbon source in liquid medium

The mix of selected bacterial isolates was cultured in 50 mL MS media in Erlenmeyer flasks with the MS media having different concentrations of crude oil: 0.5, 1, 1.5, 2, and 3. After 27 days of incubation at 30 °C, 150 rpm, then the samples after extracted, analyzed by gas chromatography (GC).[7]

## 2.7 16SrRNA gene sequences

The DNA sequence analysis method was implemented to investigate the existence of a genetic relationship between the gene encoding the 16SrRNA of bacterial isolates to compare it with the globally registered strains in the NCBI gene bank. The test was performed after obtaining the results of the PCR reaction of the gene encoding the 16SrRNA in the five isolates. The product of the PCR reaction was sent to (Macrogen/Korea), after obtaining the base sequence of the gene using the AB DNA sequencing system. The gene was analyzed by the NCBI-BLAST identity site. Finally, the isolates were recorded at the NCBI gene bank.

## 2.8. gas chromatography (GC)

Shimadzu (2014, Japan) gas chromatography used in the detection of PAHs in the laboratories of the Ministry of Science and Technology, column separation was carried out in a (30 m × 0.25 mm i.d.) DB-5 column (J&W Scientific, Folsom, CA) coated with a 0.25-μmthick film of 5% diphenyl–polydimethylsiloxane. The samples were injected in the split mode at an injection temperature of 280 °C. The column temperature was initially held at 40 °C for 1 min, raised to 120 °C at the rate of 25 °C/min, then to 160 °C at the rate of 10 °C /min, and finally to 300 °C at the rate of 5 °C/min, held at final temperature for 15 min. Detector (FID) temperature was kept at 330 °C. Helium was used as a carrier gas at a constant flow rate of 5 mL/min. GC methods are in common use because of the broad range of hydrocarbons that are detected selectively and sensitively.[8]

## 2.9. Statistical Analysis

The data were analyzed statistically according to the complete random design method to show the confirmation of bacterial strains and the concentrations and the reconciliations between them, as well as the trend analysis of the concentrations (as levels of a quantitative factor), then the differences between the averages of the factors and the combinations were compared with the Duncan multi-range method, and all statistical procedures were performed with the help of the program Ready-made statistical analysis system.[9]

## 3. Results and discussion

### 3.1 Isolation and diagnosis of bacteria

The five bacterial isolates, AM-I-1, AM-I-2, AM-I-3, AM-I-4, and AM-I-5, were diagnosed based on their phenotypic characteristics and biochemical tests, as shown in Table (1) and (2). The 16SrRNA sequence analysis was shown that the bacterial isolates were new and diagnosed for the first time, and they were registered in the GenBank under the accession numbers(LC596402, LC596403, LC596406, LC596404, LC596405).

Table (1) biochemical tests of isolates.

| No. | Tests | AM-I-1 | AM-I-2 | AM-I-3 | AM-I-4 | AM-I-5 |
|---|---|---|---|---|---|---|
| 1 | Motility | - | + | + | + | - |
| 2 | Indole | - | - | - | - | + |
| 3 | Methyl red | - | - | + | + | - |
| 4 | Voges–Proskauer | - | - | - | - | - |
| 5 | TSI | K\K | A\K | A\A | A\A | A\A |
| 6 | Oxidase | + | - | + | - | + |
| 7 | Catalase | + | + | + | + | - |
| 8 | Urease | - | - | - | - | + |
| 9 | Simmon Citrate | - | - | - | - | + |
| 10 | Coagulase | + | - | + | + | - |
| 11 | MacConkey agar | - | - | - | - | - |
| 12 | Mannitol salt agar | +\Non-F | +/F | +/Non-F | +/F | +\F |
| 13 | Blood Hemolysis | β | β | β | β | β |
| 14 | spores | - | + | + | - | - |

(+) = Positive. (-) =Negative.   β= Beta β.   A= Acid.   K = Alkaline.   F=Fermentation. Non-F=Non-Fermentation

Table (2) phenotypic characteristics Colonies and Cells

| Isolates | Nature of colonies on a solid nutrient medium | | | | | | Gram stain | Cell pool | Cell shape |
|---|---|---|---|---|---|---|---|---|---|
| | Shape | Colour | Texture | Appearance | Hight | Edge | | | |
| AM-I-1 | Circular | Yellow | Creamy | Dark | Flat | Regular | + | Tetrads | Micrococcus |
| AM-I-2 | Circular | Bright white | Creamy | Dark | Flat | Regular | + | single | Rod |
| AM-I-3 | Circular | light pink | Creamy | Dark | Flat | Regular | + | Single and binary | Rod |
| AM-I-4 | Circular | Orange-yellow | Creamy | Dark | convex | Regular | - | single | Rod |
| AM-I-5 | irregular | Light white | smooth | Dark | umbonate | undulate | + | chain | Rod |

## 3.2 The clear region of the isolates on the solid medium

Figure (1) Formation of the clear zone on solid MMS with an ethereal solution of crude oil (10% v/v) when keeping warm at 30 °C intended for 7 days.
A= AM-I-3.    B= AM-I-1.    C= Control.    D= AM-I-4.    E= AM-I-2.    F= AM-I-5.

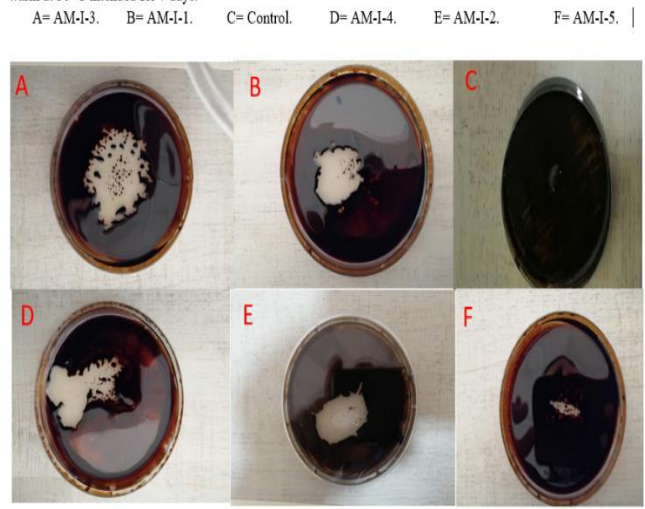

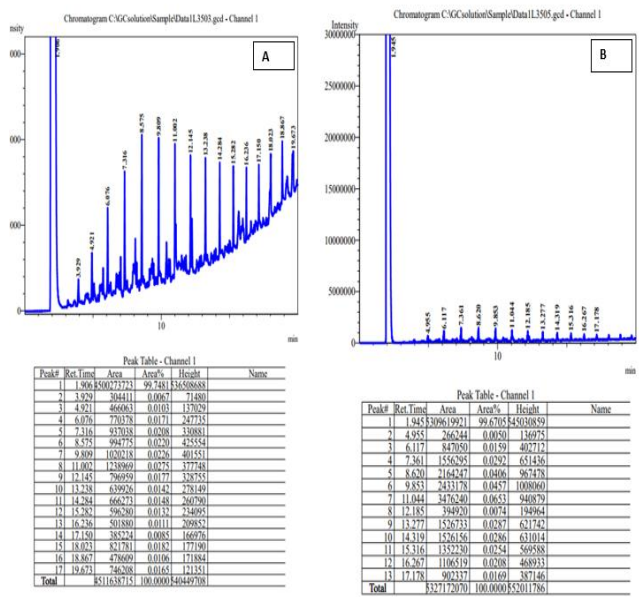

Figure (2) The GC Result [ A= Control    B= Test (Mix of the bacterial isolation)]

The growth and formation of microorganisms in the corona region around microbial colonies on a solid medium were considered as evidence of their ability to utilize crude oil hydrocarbons.[10]

The results in this study are in agreement with Latha and Kalaivani, (2012) reported that the upper bound of the halo formed around bacterial isolates cultured on a solid mineral saline medium covered with a crude oil etheric solution was observed at 30 °C for seven days. It indicated its ability to degrade crude oil hydrocarbons.[11]

## 3.3 Biodegradation of petroleum hydrocarbons

Table (3) and figure (2) indicate the results of GC, which shows the extent of degradation that occurred to the petroleum hydrocarbons, as well as the difference between the control and the mixture of bacterial isolates. The bacterial isolates possess enzymes that made them able to use crude oil hydrocarbons as a source of carbon and were adapted to live in environments contaminated with crude oil. 2% is the optimal concentration of isolates, we note that the degradation rate of aromatic hydrocarbons is: 57.32%, 69.36%, 63.71%, 75.20%, 68.60% for concentrations: 0.5, 1, 1.5, 2 and 3 respectively, and this indicates that the concentration of 2% of crude oil is the best concentration for degradation and this is what was shown by the results of GC in Table (3) and also in figure (2), as at this concentration the highest value of crude oil hydrocarbons degradation was obtained, and what was reported by (Nafl, 2020),[12] as it is with the low concentration that the enzymes that degrade are not stimulated, on the other hand, high concentrations of crude oil cause bacterial toxicity and increase the viscosity of the medium, which negatively affects the oxygen content and the pathway of hydrocarbon degradation, Perhaps the possession of microorganisms of spores enables them to thrive in exceptional environmental conditions.[13]

The results of (GC) showed that the five isolates under study are from oily decomposition bacteria. Table (3) and Fig. (3) indicate the apparent decrease in the concentrations of multiple aromatic hydrocarbons (PAHs) present in the crude oil sample, and also showed that the bacterial isolate mixture with a concentration of 2% led to a clear decomposition.

Table (3) The GC Analyze

| Name (ppm) | Control | The mix of bacterial isolates | | | | |
|---|---|---|---|---|---|---|
| | | 0.5% | 1% | 1.5% | 2% | 3% |
| Naphthalene | 1568.6 | 754.8 | 654.8 | 710.8 | 526.9 | 633.5 |
| Methylnaphthalene | 1789.4 | 569.8 | 412.9 | 523.6 | 355.8 | 406.8 |
| Acenaphthylene | 1689.6 | 745.6 | 524.8 | 589.9 | 452.9 | 541.8 |
| Fluorene | 923.9 | 412.6 | 365.9 | 400.5 | 258.9 | 374.5 |
| Phenanthrene | 1294.5 | 569.2 | 214.8 | 321.5 | 197.8 | 233.6 |
| Anthracene | 2586.8 | 956.8 | 649.8 | 745.9 | 478.9 | 674.1 |
| Fluoranthene | 1198.7 | 578.9 | 510.3 | 568.7 | 466.2 | 532.5 |
| Pyrene | 974.8 | 356.9 | 248.9 | 289.7 | 208.7 | 269.8 |
| Benz[a]anthracene | 1479.4 | 641.5 | 461.5 | 523.6 | 419.7 | 487.9 |
| Chrysene | 2189.9 | 1058.9 | 745.6 | 905.9 | 630.4 | 758.9 |
| Benzo[b]Fluoranthene | 1879.6 | 968.7 | 741.5 | 864.5 | 514.8 | 752.0 |
| Benzo[k]Fluoranthene | 869.4 | 301.5 | 258.9 | 287.9 | 198.7 | 264.1 |
| Benzo[a]pyrene | 124.8 | 51.6 | 33.6 | 41.8 | 22.9 | 36.9 |
| Dibenz[a,h]anthracene | 1478.3 | 633.6 | 418.9 | 526.9 | 369.7 | 427.9 |
| Benzo[ghi]perylene | 1393.8 | 549.8 | 326.9 | 477.9 | 214.9 | 337.9 |
| Degradation% | | 57.32% | 69.36% | 63.71% | 75.20% | 68.60% |

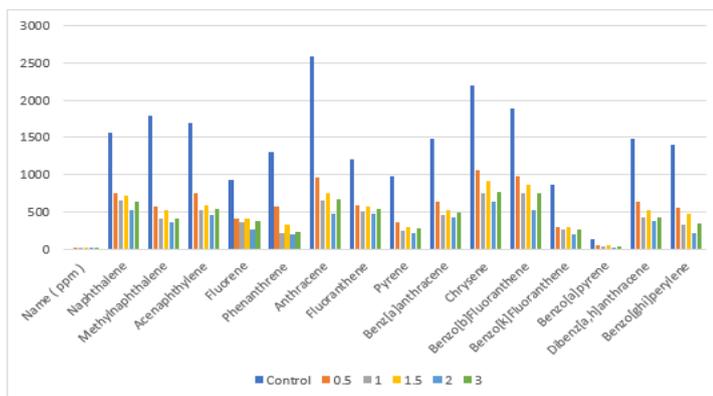

Figure (3) Diagram showing the percentage of degradation of polyaromatic petroleum hydrocarbons (PAHs) compounds.

The proportion of Naphthalene utilization at this concentration was (526.9ppm), Methylnaphthalene (355.8ppm), Acenaphthylene(452.9ppm), Fluorene(258.9ppm), Phenanthrene(197.8ppm), Anthracene(478.9ppm), Fluoranthene(466.2ppm), Pyrene(208.7ppm), Benz[a]anthracene(419.7ppm), Chrysene(630.4ppm), Benzo[b]Fluoranthene(514.8ppm), Benzo[k]Fluoranthene(198.7ppm), Benzo[a]pyrene(22.9ppm), Dibenz [a, h] anthracene(369.7ppm), Benzo[ghi]perylene(214.9ppm), a significant decrease in (polycyclic aromatic hydrocarbons) from crude oil, where the value of degradation reached 75.20%, while the percentage of degradation reached (57.32, 69.36, 63.71, 68.60) % for concentrations (0.5, 1, 1.5,3), respectively, and this indicates that all five isolates possess genes that encode enzymes that break down the crude oil compound into simple compounds to benefit from them and that these isolates have been adapted to survive in these conditions.[14]

## 4. Conclusions

In this study, new types of bacteria have been found that can break down crude oil and use aromatic hydrocarbons as a source of energy for them, and it is possible to benefit from these types in treating soils contaminated with crude oil, and the use of bacteria in a mixture at a concentration of 2% of crude oil was appropriate I give the very best percentage for oil analysis.

The conclusions section should come in this section at the end of the article, before the acknowledgments.

## 5. Conflicts of interest

There are no Conflicts to declare.

## 6. Acknowledgments

This study was conducted at the University of Mosul, College of Education for Pure Sciences, in the Laboratory of Biotechnologies in the Republic of Iraq, and special thanks go to Prof. Dr, Ibrahim Al-Hamdani, the supervisor of my thesis.